\documentclass[twocolumn,trackchanges]{aastex701}

\usepackage[version=4]{mhchem} 

\submitjournal{ApJL}

\begin{document}

\title{Oxygenated False Positive Biosignatures in Mars-like Exoplanet Atmospheres}

\author[0000-0001-7285-7925]{Margaret Turcotte Seavey}
\affiliation{Southeastern Universities Research Association, Washington, DC 20005, USA}
\affiliation{NASA Goddard Space Flight Center, Greenbelt, MD 20771, USA}
\affiliation{NASA GSFC Sellers Exoplanet Environments Collaboration}
\email[show]{margaret.turcotte@maine.edu}  

\author[0000-0003-0354-9325]{Shawn Domagal-Goldman} 
\affiliation{Mary W. Jackson NASA Headquarters, Washington, DC 20546, USA}
\email{shawn.goldman@nasa.gov}

\author[0000-0003-3099-1506]{Amber Young}
\affiliation{Mary W. Jackson NASA Headquarters, Washington, DC 20546, USA}
\email{amber.v.young@nasa.gov}

\author[0000-0003-2273-8324]{Jaime Crouse}
\affiliation{Johns Hopkins University, Department of Earth and Planetary Sciences, Baltimore, MD 21218, USA}
\affiliation{NASA Goddard Space Flight Center, Greenbelt, MD 20771, USA}
\affiliation{NASA GSFC Sellers Exoplanet Environments Collaboration}
\email{jaime.s.crouse@nasa.gov}

\author[0000-0002-0746-1980]{Jacob Lustig-Yaeger}
\affiliation{Johns Hopkins Applied Physics Laboratory, Laurel, MD 20723, USA}
\email{jacob.lustig-yaeger@jhuapl.edu}

\author[0000-0001-6285-267X]{Giada Arney}
\affiliation{NASA Goddard Space Flight Center, Greenbelt, MD 20771, USA}
\email{giada.n.arney@nasa.gov}

\begin{abstract}
\noindent
Oxygen is a well-studied biosignature. Studying potential abiotic pathways for \ce{O2} build-up in exoplanet atmospheres is essential for evaluating whether the detection of \ce{O2} would constitute a biosignature detection on other worlds. Previous modeling efforts in the literature demonstrated that detectable abiotic \ce{O2} and \ce{O3} can be produced through \ce{CO2} photolysis for rocky planets around M dwarf stars. Building on modeling approaches from previous studies, we use photochemical simulations to reassess the conditions under which \ce{O2} and \ce{O3} may accumulate through similar photochemical mechanisms. Using a Mars-like atmospheric composition and planetary parameters, we vary the hydrogen mole fraction to assess how changes in \ce{HO}x chemistry can affect the resulting accumulation of abiotic \ce{O2} and \ce{O3}. Across the range of hydrogen mole fractions explored, we obtain a maximum \ce{O2} abundance of $\sim$2.7\% for \ce{H} = 0.0065 ppm, about an order of magnitude lower than reported in the literature. This reduction is consistent with the elevated water vapor abundance adopted in our simulations, which enhances \ce{HO}x-driven recycling of \ce{CO} and \ce{O} and thereby suppresses the accumulation of \ce{O2} and \ce{O3}. Our improved understanding of how this cycle results in atmospheric false positive biosignatures is crucial towards developing future exoplanet characterization strategies.

\end{abstract}


\section{Introduction} \label{sec:intro}
M dwarf stars are the most abundant stellar type in our galaxy, which makes them a key population for studies to detect and characterize exoplanets. Population-level studies with transit and radial velocity methods show that terrestrial planets occur in roughly 50\% of such planetary systems \citep{bon13,mul15,dre15,sab21,pin22,mig25,kam25}, which make studying M dwarf systems in greater detail fundamental to understanding the variety of atmospheric compositions that could exist on such worlds. Chemical species like \ce{O2} and \ce{O3} are key diagnostic gases in the search for life because of the biological production of \ce{O2} from photosynthesis on Earth, and the consequent photochemical production of \ce{O3}. However, \ce{CO2} photolysis can be a mechanism for abiotic production of \ce{O2} and \ce{O3} allowing for substantial accumulation of both species constituting a potential false-positive for life detection. Past studies have shown that \ce{CO2}-rich atmospheres with little or no \ce{H2O} can accumulate \ce{O2} and \ce{O3} \citep{tia14,dom14,har15,hu20,ran23}. However, \citet{har18} found that \ce{O2} does not accumulate when \ce{NO} produced by lightning catalyzes the recombination of \ce{CO} and \ce{O} from \ce{CO2} photolysis.

The results of \citet{gao15} are seminal in the study of oxygen false positive biosignatures. They show that, for M dwarf terrestrial exoplanets that are \ce{CO2}-rich with a low H abundance, \ce{CO2} photolysis can produce abiotic \ce{O2} and \ce{O3} abundances comparable to modern Earth levels. Furthermore, they demonstrate that \ce{CO2} is able to be retained even with a limited water vapor abundance because catalytic cycling driven by \ce{H2O2} can serve as a source of replenishment for \ce{CO2}. Considering their model assumed a low-\ce{H2O} atmosphere and a dry surface, which decreased surface deposition rates of oxidized species, it is important to consider the role of atmospheric and surface water on the \ce{CO2} catalytic cycle for potentially habitable terrestrial exoplanets. 

Here, we explore whether \ce{O2} and \ce{O3} can be generated abiotically through similar photochemical pathways on an M dwarf terrestrial planet with Mars-like, potentially habitable conditions. In contrast to \citet{gao15}, our simulations incorporate a higher \ce{H2O} abundance, allowing us to assess how this difference influences the resulting \ce{O2} and \ce{O3} production. In Section \ref{sec:methods}, we summarize the one-dimensional photochemical model we use to produce our results, including a description of the model atmosphere and the treatment of atmospheric hydrogen. Our results for the production and stability of oxygenated species are in Section \ref{sec:results}. In Section \ref{sec:disc}, we consider the possibility of false positive detections of oxygenated biosignatures, and include an intercomparison of our results and the \citet{gao15} work. Lastly, we share our conclusions in Section \ref{sec:conc}.

\section{Methods} \label{sec:methods}
\subsection{Photochemical Model} \label{subsec:photom}
We use the one-dimensional photochemical-climate code \texttt{Atmos} for all the atmospheric simulations presented in this work \citep{arn16}. The photochemical model, which is a component of \texttt{Atmos}, uses atmospheric boundary conditions, chemical species profiles, thermodynamic parameters, and planetary parameters that are prescribed through atmospheric templates that define the model's initial state. For this study, we use the photochemical model (\texttt{PHOTOCHEM}) to calculate the abundances of species as a function of altitude for different boundary conditions. The model calculates the steady state distributions of 25 species: \ce{O}, \ce{O1D}, \ce{O2}, \ce{O3}, \ce{N}, \ce{N2}, \ce{N2O}, \ce{NO}, \ce{NO2}, \ce{NO3}, \ce{N2O5}, \ce{HNO2}, \ce{HNO3}, \ce{HO2NO2}, \ce{H}, \ce{H2}, \ce{H2O}, \ce{OH}, \ce{HO2}, \ce{H2O2}, \ce{CO}, \ce{CO2}, \ce{H2S}, \ce{SO2}, and \ce{HCl} (see Section \ref{subsec:modela}). We refer the reader to \citet{tea22} and \citet{arn16} for detailed descriptions of the \texttt{Atmos} photochemical model, and the reactions and rate constants therein.

We use a spectrum of the M dwarf GJ 436 from the MUSCLES survey \citep{fra16} for our simulations, scaled such that the total flux of the star is equal to the solar flux at 1 AU, $\sim$1360 W m\textsuperscript{-2} \citep{cla12}. We assume that the spectrum of the star is fixed throughout the duration of our simulations given that the effects of major flaring events are irrelevant on long time scales \citep{seg10,til19}. The GJ 436 spectrum features a higher FUV/NUV ratio than Sun-like stars, which causes the photolysis rates of chemical species to differ between such cases. The spectral range used in the model is 100-400 nm.

\subsection{Model Atmosphere} \label{subsec:modela}
\begin{figure}[h!]
\plotone{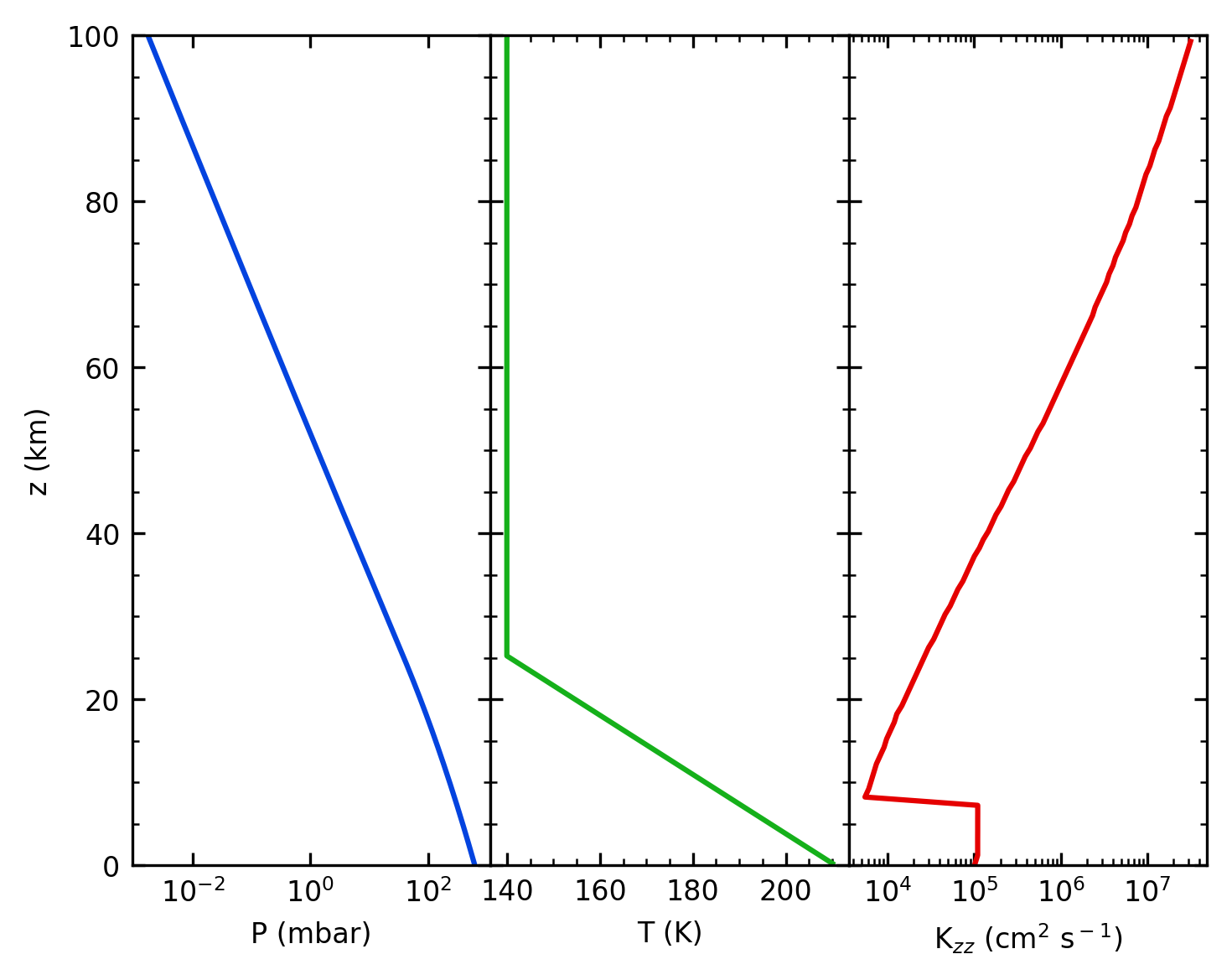}
\caption{Model atmospheric pressure \textit{P} in mbar (left), temperature \textit{T} in Kelvin (center), and eddy diffusion coefficient \textit{K\textsubscript{zz}} (cm\textsuperscript{2} s\textsuperscript{-1}) (right) as functions of altitude \textit{z} (km).
\label{fig:PTZ}}
\end{figure}

In our model, we simulate an atmosphere with a Mars-like composition and surface atmospheric pressure of 1 bar (0.987 atm) for a planet with Mars' radius and surface gravity in order to model a \ce{CO2}-dominated atmosphere. A comparison of the planetary parameters utilized in our model and those of the Caltech/JPL model used by \citet{gao15} is shown in Table \ref{tab:planet}. We use the same mixing ratios of Mars-like species as \citet{nai94} and \citet{gao15}. The pressure \textit{P}, temperature \textit{T}, and eddy diffusion coefficient \textit{K\textsubscript{zz}} profiles are shown in Figure \ref{fig:PTZ}, with the \textit{K\textsubscript{zz}} profile identical to that of \citet{gao15}. In our simulations, the model atmosphere extends from the surface to 100 km altitude, divided into layers 1 km thick.

\begin{table}[]
\caption{Comparison of Planetary Parameters}
\label{tab:planet}
\centering
\begin{tabular}{lcc}
Planetary Parameter & \texttt{Atmos} & \citet{gao15} \\ \hline
Gravity & 3.73 m s2 & 9.81 m s2 \\
Planet radius & 3.3766E6 m & 6.371E6 m \\
Solar flux scaling & 1 AU & 1 AU \\
Surface atmospheric pressure & 1 bar & 1 bar \\
Surface temperature & 240 K & 240 K \\
Troposphere height & 7.5 km & 7.5 km
\end{tabular}
\end{table}

The pressure profile is calculated assuming chemical species behave as ideal gases and that the atmosphere maintains hydrostatic equilibrium. The temperature profile is separated into the troposphere and the stratosphere. In the troposphere, the temperature profile is assumed to follow a dry adiabatic lapse rate, calculated using the temperature dependent heat capacity of \ce{CO2}, as in \citet{gao15} and \citet{mcb61}. The surface temperature is assumed to be 240 K on account of warming from atmospheric \ce{CO2} \citep{kas91}. The temperature decreases along the adiabatic profile with increasing altitude until the stratosphere and is held constant at 139 K above this point, up to the model's upper boundary at 100 km, following  \citet{nai94}. We refer the reader to \citet{gao15} for a detailed description of the eddy diffusion coefficient as a function of altitude. The temperature, pressure, and eddy diffusion coefficient profiles are held constant with time throughout the simulations.

The boundary conditions of our model firmly shape the subsequent composition of the atmosphere. Although the Mars template within \texttt{PHOTOCHEM} includes 76 chemical species, we prescribe zero-flux boundary conditions for all species except those listed in Table \ref{tab:species} to simulate comparable Mars-like atmospheres to past studies. Minor differences in atmospheric composition arise with the inclusion of constant upward fluxes for \ce{H2S}, \ce{SO2}, and \ce{HCl} in the Mars template, but the impact of those species is negligible on \ce{CO2} reactionary pathways. 

The Mars template within the model treats \ce{CO2} as an inert species, meaning the production and destruction of this species occurs at the same rate. As such, \ce{CO2} is assigned a fixed mixing ratio of 0.95 so the composition of the atmosphere is comparably rich in this species to past studies and modern Mars.

\begin{table}[]
\caption{Boundary conditions of photochemical species adopted from \citet{gao15}}
\label{tab:species}
\centering
\begin{tabular}{cc}
Species from \citet{gao15} & \texttt{Atmos} \\ \hline
\ce{O} & long-lived \\
\ce{O1D} & short-lived \\
\ce{O2} & long-lived \\
\ce{O3} & long-lived \\
\ce{O+} & --- \\
\ce{O2+} & --- \\
\ce{N} & long-lived \\
\ce{N2D} & --- \\
\ce{N2} & inert \\
\ce{N2O} & long-lived \\
\ce{NO} & long-lived \\
\ce{NO2} & long-lived \\
\ce{NO3} & long-lived \\
\ce{N2O5} & long-lived \\
\ce{HNO2} & short-lived \\
\ce{HNO3} & long-lived \\
\ce{HO2NO2} & long-lived \\
\ce{H} & long-lived \\
\ce{H2} & long-lived \\
\ce{H2O} & long-lived \\
\ce{HO2} & long-lived \\
\ce{H2O2} & long-lived \\
\ce{OH} & long-lived \\
\ce{CO} & long-lived \\
\ce{CO2} & inert \\
\ce{CO2+} & --- \\
\ce{CO2H+} & ---
\end{tabular}
\tablecomments{Long-lived species are chemically stable in the atmosphere over long timescales, short-lived species are in photochemical equilibrium such that atmospheric transport is negligible, and inert species experience equal rates of production and destruction in the atmosphere.}
\end{table}

\subsection{Atmospheric Hydrogen and Oxygen Content} \label{subsec:hyox}
Our atmospheric model differs from a desiccated Mars atmosphere with the presence of water throughout the atmospheric profile. The concentration of water is calculated along the profile with respect to pressure and temperature. For any \textit{P} and \textit{T}, there is a certain amount of water in the atmosphere equal to the relative humidity multiplied by the \ce{H2O} saturation vapor pressure. Relative humidity is the ratio of the actual amount of water vapor in the atmosphere to the maximum amount of water the atmosphere can hold at a given \textit{T}. The relative humidity of the troposphere in our model is fixed to 0.17. The \ce{H2O} saturation vapor pressure is calculated from the Clausius-Clayperon equation
\begin{equation}
\frac{\textit{d} \mathrm{ln}\textit{P}_v}{\textit{d} \mathrm{ln} \textit{T}} = \frac{\textit{m}_v\textit{L}}{\textit{RT}}
\end{equation}
since water vapor behaves like an ideal gas below $\sim$373 K \citep{kas88}. Given an isothermal temperature of the stratosphere in our model, the water content within that layer is calculated photochemically.

We vary the \ce{H2} content to study how different amounts of hydrogen depletion affect atmospheric composition. Like in past studies \citep{nai94,gao15}, we simulate six cases with total atmospheric H mole fractions \textit{f}(H)\textsubscript{tot} of (1) 26 ppm, (2) 2.6 ppm, (3) 0.27 ppm, (4) 0.032 ppm, (5) 0.0088 ppm, and (6) 0.0065 ppm. \textit{f}(H)\textsubscript{tot} is the number of hydrogen atoms in the atmosphere divided by the total number of atoms in the atmosphere. Comparatively, the H mole fraction of Earth is ${\sim}0.58$ ppm \citep{schm74}.

\section{Results} \label{sec:results}
Here we describe the resulting mixing ratio profiles for the \ce{H2} cases outlined in Section \ref{subsec:hyox}. We find that the mixing ratio profiles for most long-lived species are consistent across \textit{f}(H)\textsubscript{tot} cases. This results from the reliance on \ce{O3}, \ce{H2O2}, and \ce{H2O} by the \ce{CO2} catalytic cycle to drive replenishment of oxygenated species due to low UV flux by the M dwarf host star suppressing photolytic destruction of these species. \ce{CO2} photolysis enables the production of abiotic \ce{O2} and \ce{O3} up to 2.7\% and 1 ppm, respectively, which results in potential false positives when using \ce{O2} and \ce{O3} as keystones in the search for life on water-depleted Mars-like exoplanets.

\begin{figure*}[h!]
\plotone{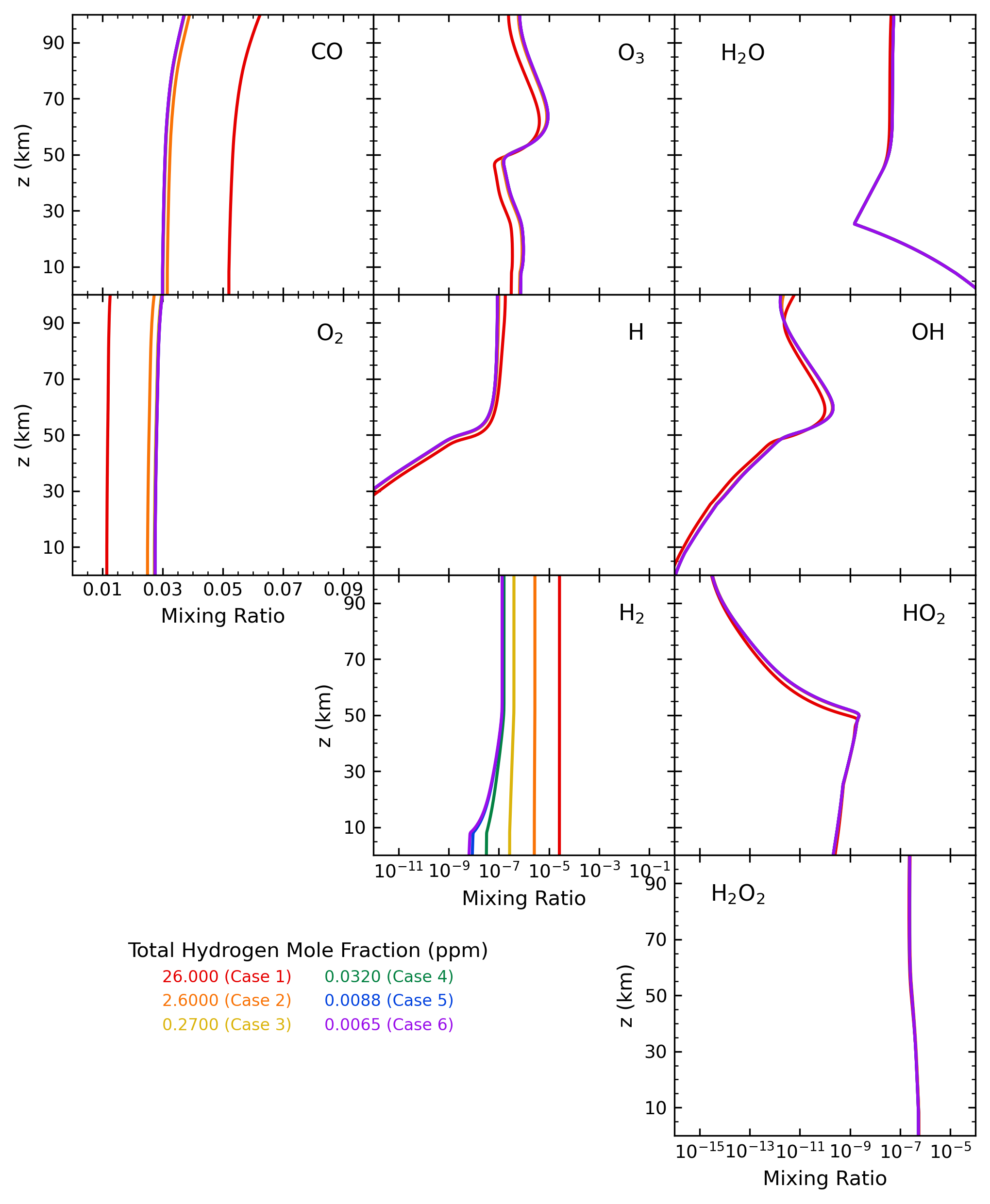}
\caption{Mixing ratio profiles for \ce{CO}, \ce{O2}, \ce{O3}, \ce{H}, \ce{H2}, \ce{H2O}, \ce{OH}, \ce{HO2}, and \ce{H2O2} for cases 1 (red), 2 (orange), 3 (yellow), 4 (green), 5 (blue), and 6 (purple) as functions of altitude \textit{z}. Note the different x-axis values for each column. The curves for cases 2, 3, 4, 5, and 6 overlap each other for most species.
\label{fig:mixrat}}
\end{figure*}

Figure \ref{fig:mixrat} shows the converged steady-state mixing ratios of key species in our model atmosphere with respect to altitude after $\sim$4 Gyr of evolution -- the age of the Solar System -- for varied atmospheric hydrogen content. Long-lived species, like \ce{CO}, \ce{O2}, \ce{H2}, and \ce{H2O2}, are well mixed below the homopause (50 km). Photolysis has a greater impact above the homopause, resulting in an increase in CO and \ce{O2} at higher altitudes. The \ce{H2} mixing ratio stabilizes due to equal rates of production and destruction. Species for which the photochemical reactions occur at lower altitudes, like \ce{H2O} and \ce{H2O2}, are well mixed above the homopause. \ce{O2} is also well mixed in the atmosphere because the mixing ratios are essentially constant with altitute for each \ce{H} mole fraction case. \ce{O3}, however, is not well mixed because of high production rates near the homopause that result in mixing ratio maxima across the \ce{H} mole fraction cases. H and OH have greatest abundances above the homopause due to removal at lower altitudes stemming from water-related chemical reactions.

\begin{figure}[h!]
\plotone{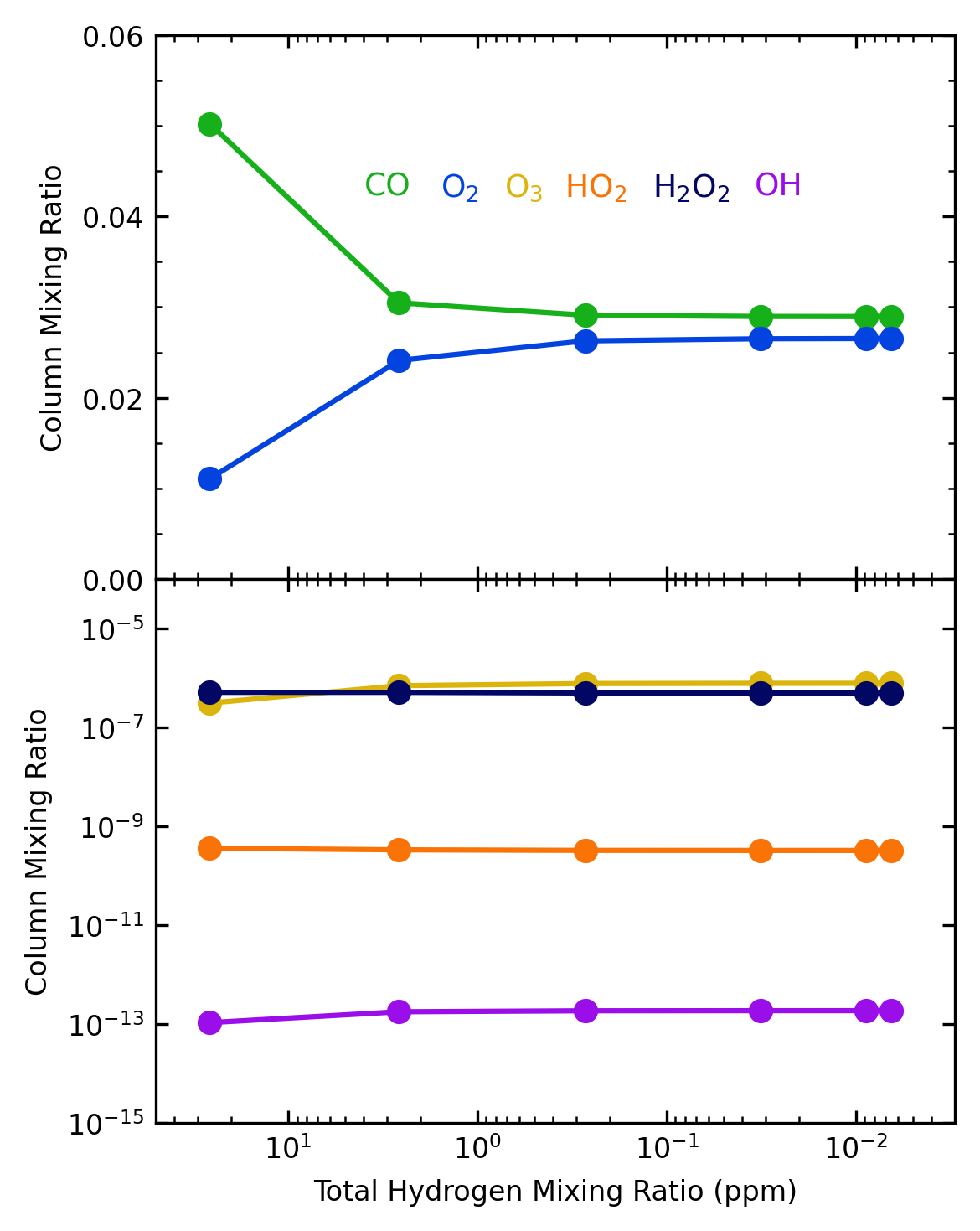}
\caption{Column-integrated mixing ratios of \ce{CO} (green), \ce{O2} (blue), \ce{O3} (yellow), \ce{HO2} (orange), \ce{H2O2} (dark blue), and \ce{OH} (purple) as functions of the total atmospheric hydrogen mole fraction for Cases 1 through 6. The results of the cases for each species are indicated by points. Note the different y-axis scales between the top and bottom panels. The mole fraction is calculated by dividing the number of hydrogen atoms in the atmosphere by the total number of atoms in the atmosphere.
\label{fig:cimr}}
\end{figure}

Trends emerge for the mixing ratios of key species as the atmospheric hydrogen content changes, as shown in Figure \ref{fig:mixrat}. The mixing ratio of \ce{CO} decreases with the \ce{H} mole fraction before leveling out at a mixing ratio threshold of about 0.03. On the other hand, the mixing ratios of \ce{O2} and \ce{O3} increase as the \ce{H} mole fraction decreases. Some species, notably \ce{HO2} and \ce{H2O2}, experience little variation as the hydrogen content changes. Figure \ref{fig:cimr}, which details the column-integrated mixing ratios of key species as a function of atmospheric hydrogen content, clearly shows this trend. 

\section{Discussion} \label{sec:disc}
\subsection{False Positive Biosignatures} \label{subsec:fals}
Our results show that mixing ratios for \ce{O2} of $\sim$2.7\% and $\sim$1 ppm for \ce{O3} are possible due to photochemistry for terrestrial exoplanets around M dwarf stars. These values are 10 times smaller than on modern Earth \citep{yun99}, where \ce{O2} and \ce{O3} abundances are produced via photosynthesis. In this section, we explore the photochemical reactions driving \ce{O2} and \ce{O3} production on wet Mars-like exoplanets and consider the potential for false positive biosignature detections with future observatories. 

To simplify the processes of our model atmosphere, we emulate past studies to assume zero-flux boundary conditions for all neutral species \citep{nai94,gao15}. This enables our focus on the effects of UV flux, water content, and \ce{CO2} content on the production of \ce{O2} and \ce{O3} false positive biosignatures. Maintaining a zero-flux boundary condition is the same as assuming all fluxes are non-zero and cancel each other out. As \ce{H2} decreases from cases 1 to 6, the mixing ratios of \ce{O2} and \ce{O3} increase due to water vapor in the model atmosphere. The \ce{O2} and \ce{O3} content within the model atmosphere are highly stable in the troposphere and stratosphere. The production of \ce{O2} through the reaction of \ce{O} radicals is a factor of 2 slower than the destruction of \ce{O2} via photosynthesis. Some atmospheric escape of oxygen radicals to space inhibits the accumulation of an oxygen sink that would otherwise fuel further \ce{O2} production. The abundances of \ce{O3} in the atmosphere are stable across all H mol fraction cases because of near-equal rates of production (from \ce{O2} and \ce{O}) and destruction through photolysis. The behavior of abiotic \ce{O2} and \ce{O3} in our atmospheres is consistent with past studies \citep[see][]{seg07,gao15}. However, the concentrations of these species differ due to differences in characteristics of the host planet (see Section \ref{subsec:imco} for more details).

\ce{H2O} catalyzes the production of \ce{O2} and \ce{O3}, which typically works against \ce{CO2} photolysis. Across all H mol fraction cases in our atmosphere, \ce{H2O} is destroyed at a rate $\sim$10\textsuperscript{4} greater than its production, which creates an OH reservoir that reacts with CO to sustain the zero-flux conditions of \ce{CO2} and \ce{H2}. High \ce{H2O2} mixing ratios across the different cases play a role in the stability of \ce{CO2} in the atmosphere through reactions with OH. The rate of \ce{CO2} photolysis is 10\textsuperscript{2} to 10\textsuperscript{3} times greater than the \ce{CO2} production across all cases, which results in detectable CO and O that contribute to subsequent reactions in the catalytic cycle to form \ce{O2} and \ce{O3} false positive biosignatures.

\subsection{Inter-model Comparison} \label{subsec:imco}
The results of our study using a modified version of the \texttt{Atmos} one-dimensional \texttt{PHOTOCHEM} model are consistent with those of \citet{gao15} using the Caltech/JPL one-dimensional photochemical model \citep{all81}, as modified by \citet{nai94} for Mars-like atmospheres. There are identical trends for the mixing ratios of H because of the fixed mixing ratio of \ce{H2}. The behavior of \ce{O2} is consistent in the troposphere, although the mixing ratio values differ by nearly a factor of 10. Smaller reaction rates for \ce{O3} in the stratosphere can be seen in both models as the concentration decreases with increasing altitude above the homopause.

Despite similarities between our results and those of \citet{gao15} (see Table \ref{tab:species}), there are differences between our models that influence results. Model limitations with \texttt{PHOTOCHEM} cause differences in planetary parameters as shown in Table \ref{tab:planet}, which prevent this study from being a true inter-model comparison. Our model utilizes a Mars-size planet with Mars' gravity, whereas \citet{gao15} used an Earth-sized planet with Earth's gravity. These discrepancies lead the influence of gravity to be nearly 3 times greater in their model, increasing the reaction rates for all species near the planetary surface within the troposphere and creating a larger reservoir in which potential false positive biosignatures can accumulate.

\texttt{PHOTOCHEM} treats \ce{CO2} as an inert species as well, which prevents us from gleaning insights on the interactions between \ce{CO2} and solar UV at different altitudes and with varying \textit{f}(H)\textsubscript{tot}. Although the trends of different species are consistent, discrepancies in planetary parameters and water content (as described in Section \ref{subsec:fals}) cause differences in mixing ratio values for different species. \ce{CO} decreases with decreasing \textit{f}(H)\textsubscript{tot} in \texttt{Atmos} because of the inert treatment of \ce{CO2} and higher water content. \ce{H2O} is over 10\textsuperscript{4} times greater at the surface in our model, which shows a sufficient surface reservoir for the accumulation of this important habitability indicator.

The consistency of mixing ratio values across the range of \textit{f}(H)\textsubscript{tot} in our model strongly indicates that the \ce{CO2} catalytic cycle is influenced by the presence of \ce{H2O} in the atmosphere. This shows that the water content can be a potential indicator of both biotic and abiotic worlds depending on if the drivers of atmospheric chemistry are photolytic or photosynthetic in nature.

\section{Conclusion} \label{sec:conc}
Our work explored the potential for false positive detections of \ce{O2} and \ce{O3} as an atmospheric biosignature on a \ce{CO2}-rich terrestrial exoplanet with elevated water vapor concentrations around an M dwarf star within the context of past photochemical oxygen false positive studies, like \citet{gao15}. We evaluated the stability of these atmospheres with respect to the production and retention of oxygenated species, and found that these atmospheres are stable because of reactions related to the \ce{CO2} catalytic cycle. The treatment of \ce{CO2} as an inert species in our model preserved it within the atmosphere, which allowed a closer inspection of the underlying reactions driving the production of \ce{O2} and \ce{O3} and the influence of \ce{H2O}.

Although the \ce{O2} and \ce{O3} mixing ratio values produced by our model are $\sim$10 times smaller than those produced by biological photosynthetic organisms on modern Earth, the trends are comparable to past studies (e.g., \citet{gao15}). The mixing ratio profiles differed due to our assumption of a low-\ce{H2O} atmosphere with a depleted surface, which increased the surface deposition rates of oxidized species. These atmospheric conditions may be applicable to observational characterization of terrestrial exoplanets orbiting in the habitable zones of M dwarfs \citep{lug15}. Because high \ce{O2} and \ce{O3} mixing ratios may be observationally detectable, understanding how an \ce{H2O}-depleted but not truly desiccated planetary environment influences these abundances can differentiate between a potentially habitable world with biologically-influenced atmospheric disequilibria from a water-depleted Mars-like world with atmospheric disequilibria stemming from photochemistry.

We were model-limited in this study because of the inert treatment of \ce{CO2}, which prevented elucidation of the relationship between \ce{CO2} and solar UV for \ce{H2O}-depleted terrestrial exoplanets. We faced challenges in conducting a true inter-model comparison through differences in planetary parameters, although our results are conclusive for planets of Martian size and surface gravity.

Future research centered on exploring the relationship between these \ce{CO2}-rich, \ce{H2O}-poor terrestrial exoplanets and their host stars across a range of M dwarfs, notably as it relates to stellar activity, has the potential to elucidate further insights related to the production of false positive oxygenated species. Comparative modeling studies are crucial towards better understanding atmospheric biosignatures, their false positives, and the underlying chemical processes that drive their accumulation. A better understanding of potential abiotic biosignature detections from this study and others drive effective differentiation between desiccated and potentially habitable worlds with future observational missions.

\begin{acknowledgments}
This material is based upon work supported by NASA under award numbers 80GSFC21M0002 and 80GSFC24M0006 through the Center for Research and Exploration in Space Sciences (CRESST-II). A.V.Y. would like to acknowledge support from the GSFC Sellers Exoplanet Environments Collaboration (SEEC), which is supported by NASA's Planetary Science Division's Research Program.
\end{acknowledgments}

\begin{contribution}
M.T.S. performed the modeling, analysis, and data visualizations; led data interpretation; and wrote and submitted the manuscript. S.D.G. supervised the investigation and contributed to data interpretation and analysis. M.T.S. and S.D.G. conceptualized the research and developed the methodology. A.V.Y. provided software resources for data visualization. A.V.Y., J.C., and J.L.Y. contributed to modeling, plotting, and interpretation of data; and conducted supervision. G.A. contributed to data interpretation. All authors contributed to manuscript revision.
\end{contribution}

\bibliography{falseoxygen.bib}
\bibliographystyle{aasjournalv7}

\end{document}